\documentclass[aps]{revtex4}  
\usepackage{graphicx}
\usepackage{dcolumn}
\usepackage{bm}
\usepackage{rotate}  
\usepackage{epsfig}  
\newcommand \be  {\begin{equation}}   
\newcommand \bea {\begin{eqnarray} \nonumber }   
\newcommand \ee  {\end{equation}}   
\newcommand \eea {\end{eqnarray}}   
    
\begin{document}   
    
 
   \title{Non linear susceptibility in glassy systems: a probe for  
   cooperative dynamical length scales}   
  
\author{Jean-Philippe Bouchaud$^{+}$}   
\email{bouchau@spec.saclay.cea.fr}   
\author{Giulio Biroli} 
\email{biroli@spht.saclay.cea.fr}   
\affiliation{$^{+}$Service de Physique de l'{\'E}tat Condens{\'e},  
Orme des Merisiers,    
CEA Saclay, 91191 Gif sur Yvette Cedex, France, and\\   
Science \& Finance, Capital Fund Management, 6-8 Bd  
Haussmann, 75009 Paris, France.\\   
$^{\dagger}$Service de Physique Th{\'e}orique, Orme des Merisiers,  
CEA Saclay,    
91191 Gif sur Yvette Cedex, France.   
} 
   
 \date{\today}   
 \begin{abstract}   
 We argue that for generic systems close to a critical point, an  
 extended Fluctuation-Dissipation relation connects the low frequency    
 non-linear (cubic) susceptibility to the four-point correlation  
 function. In glassy systems, the latter contains    
 interesting information on the heterogeneity and cooperativity of  
 the dynamics. Our result suggests that if the abrupt   
 slowing down of glassy materials is indeed accompanied by the  
 growth of a cooperative length $\ell$, then the non-linear,    
 $3 \omega$ response to an oscillating field (at frequency $\omega$) 
should substantially  
 increase and give direct information on the   
 temperature (or density) dependence of $\ell$. The analysis of the  
 non-linear compressibility or the dielectric   
 susceptibility in supercooled liquids, or the non-linear magnetic  
 susceptibility in spin-glasses, should give access to  
 a cooperative length scale, that grows as the temperature is  
 decreased or as the age of the system increases. Our   
 theoretical analysis holds exactly within the  
 Mode-Coupling Theory of glasses. 
 \end{abstract}  
 \pacs{Valid PACS appear here}
 \maketitle   
  
 \section{Introduction}   
  
 A yet unexplained property of fragile glasses is the extremely fast  
 rise of their relaxation   
 time (or viscosity) as the temperature is lowered, much faster than  
 predicted by a simple    
 thermal activation formula \cite{Nature}. If  
 interpreted in terms of an effective activation energy,   
 the latter increases by a factor five to ten between $1.5 \, T_g$  
 and the glass transition temperature $T_g$.    
 The basic mechanism for this increase is not well understood, but  
 it is reasonable to think that it is intimately   
 related to cooperative effects \cite{AG,Debenedettibook} and possibly to the  
 presence of an underlying critical point    
 \cite{KTW,Sethna,Tarjus,MP,FP,BB1,BB2,WBG}. The dynamics becomes  
 sluggish and the activation energy increases because   
 larger and larger regions of the   
 material have to move in a correlated way to allow for a  
 substantial motion of individual particles.   
 Long {\it time} scales   
 must be somehow associated with large {\it length} scales. Although  
 the idea of a cooperative length has been discussed in the context  
 of glasses for many years \cite{AG,Ediger}, it is only    
 recently that   
 proper measures of cooperativity (and of the size of the  
 rearranging regions) were proposed theoretically \cite{FP} (see
 \cite{Harrowell} for earlier insights) and   
 measured in numerical simulations  
 \cite{Harrowell,Onuki,Glotzer,Berthier1} (see also  
 \cite{Ediger,Weeks,Dauchot} for related   
 experimental work). The idea is to measure how the dynamics is  
 correlated in space; technically, this involves a   
 four-point correlation function which measures the   
 spatial correlations of the temporal correlation 
 (see Eq. (\ref{chi4}) below for a  
 more precise definition). 
 Recent extensive numerical evaluations of this four-point  
 correlation function in Lennard-Jones systems have    
 confirmed the existence of a growing length scale as temperature is  
 decreased \cite{Glotzer,Berthier1,WBG}, and have shown that   
 different observables, such as the relaxation time or the diffusion  
 constant, scale as powers of this length,  
 emphasizing its crucial importance as far as the physics is  
 concerned. In the framework of granular systems, diverging 
 length scales near the jamming transition have also been reported 
 in numerical studies of model systems \cite{Silbert}.
  
 Although many different theoretical approaches to the glass  
 transition \cite{KTW,Sethna,Tarjus,MP,FP,BB1,BB2,WBG}   
 can potentially explain the existence of such a growing dynamical  
 correlation length, these theories lead to rather   
 different quantitative predictions for the behaviour of the  
 four-point correlation function (see \cite{TWBBB}).   
 Thus, experiments measuring directly this four-point function would  
 be extremely valuable to refine our understanding of the glass phenomenon and  
 prune down the number of candidate models.   
 Up to now, unfortunately, only indirect experimental indications of a  
 cooperative length scale associated to heterogeneous   
 dynamics have been reported \cite{Ediger,Weeks}.   
  
 On a different front, that of spin-glasses, length scale ideas have  
 also been expressed in the recent years to account   
 for non equilibrium phenomena such as aging, memory and  
 rejuvenation effects \cite{FH,BM,KH,Bou,Ghost,LudoKT}.   
 Although spin-glass order is not easy to define nor to detect, the  
 idea is that some kind of domain growth occurs,   
 whereby spin-glass correlations establish on larger and larger  
 length scales as the age of the system increases.   
 The growth of this ``coherence length'' has been established  
 numerically by comparing two replicas of the same   
 system \cite{Rieger,Takayama,Marinari,BerB,BerY}. This trick is  
 obviously inaccessible to experimentalists, who   
 have nevertheless provided indirect evidence of a growing length  
 scale, and some indications on its rate of growth with    
 time and temperature \cite{Joh,Bou,Yosh-Swed,Dup}. Again, a direct measure of  
 this length scale is lacking -- finding a clear-cut experimental    
 signal of a cooperative length in disordered, amorphous systems  
 would certainly be a major breakthrough \cite{Ediger}. 
  
 The aim of this paper is to point out that in {\it slow} glassy  
 systems at equilibrium, the non-linear (cubic)   
 response to an external field (electric, magnetic, pressure, etc.)  
 in fact probes directly the four-point correlation   
 function mentioned above, and therefore the cooperative length it  
 may contain. Our main prediction, detailed below, is that   
 the $3\omega$ harmonic response to an a.c. field of frequency  
 $\omega$ and amplitude $h$ is given by   
 $\chi_3(\omega,T)\, h^3$, where the non-linear susceptibility  
 $\chi_3$ behaves at low frequency as:   
 \be\label{result}   
 \chi_3(\omega,T) = \frac{\chi_s^2}{k_B T} \ell^{2 -  
 \overline{\eta}} {\cal H}(\omega \tau).   
 \ee   
 In the above relation, $\chi_s$ is the static linear  
 susceptibility, ${\cal H}$ a certain complex function that   
 depends weakly on temperature, and $\tau$ is the temperature  
 dependent relaxation time of the system,   
 which can be directly measured using the linear susceptibility. The  
 cooperative length $\ell$ (measured in units of the microscopic  
 length $\xi$ obtained from the point correlation function) is expected to   
 grow as the temperature is reduced, and $\overline{\eta}$ an  
 exponent    
 related to the spatial structure of the four-point correlation  
 function \footnote{Note that the ratio of non-linear   
 effects to linear effects is, in order of magnitude, given by  
 $\chi_s h^2 /k_B T$, i.e. the ratio of the   
 energy of the field to the thermal energy.}.  
 The above prediction holds for equilibrium systems; we will furthermore see
 below that in the case of glasses and spin-glasses in a field, ${\cal H}(0)=0$.  
 Below the glass transition temperature, on the other hand, the  
 system by definition falls out of equilibrium. Its dynamics   
 becomes non stationary and exhibits aging, which means that the  
 effective relaxation time of the system increases with the   
 age $t_w$ of the system \cite{Sitges,Review}. This increase of the  
 relaxation time is again most probably related    
 to the growth of a coherence length in the system, $\ell_w =  
 \ell(t_w)$. Assuming simple aging behaviour, the 
 generalization of the equilibrium result (\ref{result}) then reads:  
 \be\label{resultw}   
 \chi_3(\omega,t_w) =  \frac{\chi_s^2}{k_B T}  
 \ell_w^{2-\overline{\eta}} {\cal F}(\omega t_w),   
 \ee   
 which should allow one to extract from non-linear aging susceptibilities  
 a non equilibrium coherence length, in a    
 much more direct way than previous attempts. (In the above  
 equation, $\cal F$ is another scaling function, which    
 also contains possible violations of the standard Fluctuation  
 Dissipation Theorem and the appearance of a non trivial, $t_w$ dependent,  
 effective temperature \cite{Leticia}).    
 Our central results, Eqs. (\ref{result},\ref{resultw}), that we  
 will motivate below, states that (a)    
 the non-linear susceptibility has the same frequency scaling as the  
 linear susceptibility, which is 
 not surprising and (b) it grows as the cooperative length  
 increases, which   
 should allow a direct experimental test of the relationship between  
 length and time scales in glassy systems.  
  
 As for comparison with previous works, the    
 divergence of the {\it static} non-linear   
 susceptibility at the spin-glass (at zero field) or dipolar-glass transition, displayed   
 by Eq. (\ref{result}) at $\omega=0$, is of course  
 well documented, both theoretically   
 \cite{SG} and experimentally \cite{SG,Vincent,Maglione}. The  
 generalization to the dynamical non-linear susceptibility  
 in the critical region was also discussed  
 \cite{Zippelius,FH,ParisiRanieri} but not, to the best of our  
 knowledge, its generalization to the non-equilibrium, aging regime, Eq. (\ref{resultw}).    
 The situation for glass-forming liquids is quite different, since  
 no static phase transition with a diverging   
 static susceptibility has ever been identified, neither in  
 experiments nor in simulations. 
 Purely based on an analogy with spin-glasses, it was suggested in  
 \cite{Wu} that the non-linear dielectric constant   
 of molecular glasses might grow as the glass phase is approached  
 (although this was not borne out by the experiments   
 done at that time \cite{Wu}). A similar suggestion was made in  
 \cite{Caruzzo} concerning the non-linear compressibility  
 of soft sphere binary mixtures,   
 with numerical results that are not incompatible with a substantial  
 increase of $\chi_3(\omega \sim 0)$ as the temperature is   
 lowered. We will show below that such a growth is indeed expected,  
 although the theoretical situation for glass formers   
 is much less clear than for spin-glasses -- in particular, $\chi_3$  
 although is growing may never diverge in glass formers. Different scenarii   
 for the glass transition can be envisaged and lead to quite  
 different predictions, for example on the value of   
 $\overline{\eta}$ and on the relationship between $\ell$ and $\tau$  
 (or $t_w$).   
  
 In the following section we will give some physical arguments that  
 motivate our results, and muster the predictions  
 of different theoretical models for glass-formers. A more detailed  
 and technical discussion is then presented in   
 Section III. Finally our conclusions are presented in Section IV.   
  
 \section{Physical arguments and results}    
  
 \subsection{Spin-glasses}  
  
 \subsubsection{Order parameter and non-linear susceptibility}  
  
 Let us first focus on spin-glasses in zero external magnetic field, $H=0$. These systems are known, 
 both theoretically and experimentally,   
 to have a non zero transition temperature below which the  
 magnetization profile, $\langle s_x \rangle$,    
 freezes into one (or more) amorphous configurations. The ordered  
 state is characterized by a non zero Edwards-Anderson (EA)  
 parameter $q=[\langle s_x \rangle^2]$, where $\langle ... \rangle$  
 indicates thermal averaging and the brackets a   
 spatial (or disorder) average. These systems display an unusual type  
 of long-range order, which cannot be detected using   
 either one body or two-body spin-spin correlations: because the  
 ordering pattern is random, the average magnetization   
 $[\langle s_x \rangle]$ remains zero and the spin-spin correlation  
 $[\langle s_x s_y\rangle]$ short-ranged, even in   
 the spin-glass phase. Correspondingly, the linear susceptibility,  
 related by a fluctuation dissipation theorem ({\sc fdt}) to the integral of the spin-spin   
 correlation function, does not diverge as $T_g$ is approached, even  
 if some long-ranged correlations appear in the   
 system. The way to get rid of the spurious cancellation between   
 strongly correlated and strongly anti-correlated spins is well  
 known: exactly as one should square $\langle s_x \rangle$   
 to obtain a non zero Edwards-Anderson parameter, one should also  
 square $\langle s_x s_y\rangle$ before  
 averaging over disorder. The integral over space of that quantity  
 now diverges as $T_g$ is approached, and  
 in fact has two interesting physical interpretations. The first one  
 is the susceptibility of the spin-glass order parameter  
 to small random ordering fields. Imagine one adds small random  
 magnetic fields $h_x$ on every site. Using   
 linear response, one can write, for a given  
 sample at $T > T_g$ and $H=0$:  
 \be  
 \delta \langle s_x \rangle = \frac{1}{k_B T} \sum_y \langle s_x  
 s_y\rangle_{0} h_y,  
 \ee  
 where the subscript $0$ means that the correlation functions are 
 evaluated at zero external field. 
 Squaring this relation, summing over $x$ and averaging over the  
 random fields   
 gives the sensitivity of the EA order parameter to a random pinning  
 field:  
 \be  
 \chi_{SG} = \frac{\partial q}{\partial \langle h^2 \rangle} =  
 \frac{1}{N(k_BT)^2}  \sum_{x,y} [\langle s_x s_y\rangle_{0}^2].   
 \ee  
 Clearly, the divergence of $\chi_{SG}$ signals an incipient  
 instability towards an ordering pattern favoured by   
 the small pinning fields, exactly as the divergence of the usual  
 two-body susceptibility signals an instability   
 to ferromagnetic order, triggered by a small (uniform) magnetic  
 field.  
 
 As defined above, $\chi_{SG}$ has a clear  
 theoretical interpretation but seems hard to access experimentally.  
 Fortunately, there is a direct relation between  
 $\chi_{SG}$ and the non-linear susceptibility, which can be  
 directly measured. The intuitive idea is that the   
 non-linear susceptibility is actually a measure of the (quadratic)  
 dependence of the linear susceptibility on the external   
 field. Using {\sc fdt} the change of the connected  
 correlation function between two spins (and hence of the   
 linear susceptibility) induced by the field contains the term:  
 \be\label{chi3sg}  
 \delta [\langle s_x \rangle \langle s_y\rangle] \simeq   
 \sum_{z,z'} \langle s_x s_z \rangle_{0} \langle s_y s_{z'}  
 \rangle_{0} \frac{h^2}{(k_B T)^2}  
 \ee  
 Averaging over space (or over disorder), only the terms ($z=y,x$,  
 $z'=x,y$) survive, the first one giving   
 $[\langle s_x s_y\rangle_{0}^2]$ as in $\chi_{SG}$. A more precise  
 treatment for Ising spins at zero field leads to the   
 exact relation $\chi_3(\omega=0) = - (3\chi_{SG}-2)/(k_B T)^3$.  
 Therefore, the static non-linear susceptibility   
 of spin-glasses diverges at the spin-glass transition temperature,  
 a well-known effect that allows one to measure some   
 of the critical exponents experimentally \cite{SG,Vincent}. The physics  
 behind the correlation   
 induced amplification of $\chi_3$ is clear: the influence of the  
 polarization of spin $s_x$ on $s_y$ may be either   
 positive or negative, but it has the  
 same sign as the reverse influence of $s_y$ on $s_x$. Therefore,  
 the quadratic effect of an external field $h$  
 on the dynamical correlation between any pair of spins has a well  
 defined sign, in turn leading to a diverging  
 non-linear susceptibility as the size of correlated regions  
 increases, even if the linear susceptibility remains small. 
 
 \subsubsection{Non zero external field}   
 
 The case where a non zero external field $H$ is present is more subtle. In mean-field, 
 the spin-glass phase survives in a whole region of the $T,H$ plane, below the de Almeida-Thouless ({\sc at})
 line. The spin-spin correlation function $[(\langle s_x s_y\rangle - 
 \langle s_x \rangle \langle s_y\rangle)^2]$ is long-ranged in the whole spin-glass phase but 
 is no longer directly related to the static non-linear susceptibility. Some exact compensation 
 mechanism \cite{DGMMZ,Young} actually cancels the divergence in the combination of four-spin correlations 
 appearing in $\chi_3(\omega=0)$. Therefore, the non-linear susceptibility is 
 finite in the whole spin-glass phase. There is in particular no divergence of $\chi_3(\omega=0)$ on the {\sc at} line,
 except at $H=0$; rather, the non-linear susceptibility is discontinuous across the {\sc at} phase transition 
 \cite{Crisanti}. Within the droplet theory, on the other hand, the spin-glass is detroyed by any non zero field; 
 both the spin-glass and non-linear susceptibilities are finite when $H \neq 0$ \cite{Bray,FH}. For $H = 0$, a compensation 
 mechanism similar to that of mean-field glasses is also at play, but does not prevent the non-linear susceptibility
 to diverge for all $T < T_g$ \cite{FH}.
 
 \subsubsection{Dynamical non-linear susceptibility}  
  
 The above qualitative arguments for the static non-linear susceptibility 
 can be extended to the dynamical case as well. As will be recalled below, the   
 dynamical {\sc fdt} gives:  
 \be  
 \langle s_x \rangle(t_1) = \frac{1}{k_B T} \sum_y \int dt_3  
 \frac{d}{dt_3}\langle s_x(t_1) s_y(t_3)\rangle_{0} h_y(t_3).  
 \ee   
 Therefore, the change in the connected dynamical correlations between  
 $s_x(t_1)$ and $s_y(t_2)$, induced by a uniform, but   
 time dependent external field, will contain a term like:  
 \be  
 \sum_{z,z'} \int dt_3 dt_4   
 \frac{d}{dt_3}\langle s_x(t_1) s_z(t_3)\rangle_{0}  
 \frac{d}{dt_4}\langle s_y(t_2) s_{z'} (t_4)\rangle_{0} h(t_3)  
 h(t_4).  
 \ee  
 Repeating the same argument developed in the static case, i.e.  
 averaging over space (or disorder) and using {\sc fdt} to relate  
 the connected correlation function to the dynamical linear susceptibility 
 leads to a non-linear response function that reads:  
 \be  
 \chi_3(t_1;t_2,t_3,t_4) \sim \sum_y  \frac{d^3}{dt_2 dt_3 dt_4}  
 [\langle s_x(t_1) s_y(t_3)\rangle_{0}  
 \langle s_x(t_4) s_y(t_2)\rangle_{0}].  
 \ee  
 Taking $t_1,t_2,t_3,t_4$ all within an interval of the order of the  
 relaxation time $\tau$ of the   
 system, we see that the correlation function entering $\chi_3$  
 above defines a cooperative length scale $\ell$,  
 such that the dynamics of $s_x$ and $s_y$ within this time interval  
 is dominated by common events. This in  
 turn leads to our scaling prediction, Eq. (\ref{result}), near the  
 transition temperature. The exact result   
 for the dynamical $\chi_3(t_1;t_2,t_3,t_4)$ needs to be worked out  
 carefully (see Section III), since {\sc fdt} for higher order correlations is 
 more involved than for two point functions \cite{Guilhem}.   
 Although different from the above naive expression, it indeed  
 contains four-spin correlation functions that   
 capture the cooperativity of the dynamics.   
 Intuitively, again, the non-linear response is strong if on the  
 scale of the relaxation time, two spins feedback on   
 each other's dynamics -- this cross correlation is squared and  
 survives averaging, even if the correlation itself  
 is of random sign.   
  
 \subsubsection{The aging regime}   
  
 In the low temperature, spin-glass phase, the relaxation time  
 $\tau$ is infinite and the age of the system $t_w$  
 plays a crucial role -- all time dependent correlation functions  
 depend explicitly on $t_w$ \cite{Review}. However,  
 exciting the system with a field of frequency $1/t_w$ will give the  
 non-linear response of a spin-glass equilibrated only up to a   
 certain length scale $\ell_w = \ell(t_w)$ \footnote{Within a field  
 theoretical perspective \cite{ChamonCugliandolo}   
 the growing correlation length $\ell_w$ is   
 due to soft-modes, related to the reparametrization invariance of  
 the dynamical action.}. Interestingly, contrarily to standard ferromagnets,   
 spin-glasses are thought to be {\it critical} in their  
 whole low temperature phase, in the sense that the space integral  
 of the connected correlation function $[\langle s_x s_y\rangle_{c}^2]$ diverges for all $T < T_g$.
 Within the mean-field replica theory, the static non-linear susceptibility within one phase is, as mentioned 
 above, finite (except for $T=T_g$ and $H=0$) \cite{DGMMZ}. We however think that the mechanism cancelling the
 divergence does not operate at finite frequencies, and that an equation similar Eq. (\ref{resultw}) will hold in the 
 aging phase, but with an infinite number of time domains rather than the simple scaling variable $\omega t_w$ 
 \cite{Sitges,Review}. The explicit calculation of ${\cal F}$ in the context of a spherical p-spin model
 would be extremely interesting; in particular one may ask whether the effective temperature appearing in the
 non-linear response is the same as that appearing in the linear response \cite{ABB}. 
 
 In the droplet picture \cite{FH,Bray}, on the other hand, the static non-linear susceptibility 
 diverges for all $T < T_g$, provided $H=0$, and one should certainly observe a non-linear 
 susceptibility increasing as in Eq. (\ref{resultw}), although the original droplet model 
 with activated dynamics would rather predict a function ${\cal F}$ of $\ln \omega/\ln t_w$ 
 and a logarithmic growth of $\ell(t_w)$. The peak value $\chi_3(t_w,\omega=1/t_w)$ should grow
 as $\ell(t_w)^{2-\overline{\eta}}$. The numerical value of $2-\overline{\eta}$ 
 is yet unknown, but following Fisher and Huse \cite{FH}, one may expect $d - 3 \theta \leq 2-\overline{\eta} \leq 
 d - \theta$ with $\theta \approx 0.2$ in three dimensions. 
  
\subsection{Structural glasses}  
  
\subsubsection{Four-point density functions}  
  
Let us now discuss structural glasses. The important lesson we  
learn from spin-glasses is  
that a non-trivial amorphous type of long-range order can set in.  
In the case of glasses, the subtlety  
comes from the absence of quenched disorder; however, there has now  
been many papers exploring the idea  
of self-induced disorder which could drive a similar transition in  
homogeneous, frustrated systems (see, e.g. \cite{Review} and refs. therein).   
This has led, in particular, to the ``Random First Order  
Transition'' scenario \cite{KTW}, where a glass transition   
of the same nature as the spin-glass transition in mean-field  
$p$-spin models takes place (see \cite{MP} for recent quantitative results). 
Whether a true transition of this type can exist in real 
systems with finite range  
interactions is still an actively  
debated issue; it is nevertheless extremely fruitful to explore the  
properties of systems for which   
this transition is, in some sense, nearly present. The order  
parameter in the would-be glass phase is the   
amplitude of the frozen in (random) density fluctuations $\delta  
\rho_x$. As for spin-glasses, the average of  
this quantity is zero, but $\langle \delta \rho_x \rangle^2$ is  
not, and plays the role of the Edwards-Anderson  
parameter. Similarly, one expects $\langle \delta \rho_x  \delta  
\rho_y \rangle$ not to show any interesting  
features (beyond that typical of a liquid structure factor),  
whereas its square may reveal long-range cooperative  
dynamics. The analog of the spin-glass and non-linear  
susceptibilities discussed previously can be easily found in the case of  
glasses: the former can be seen as the susceptibility to a random  
external pinning field \cite{Monasson} that triggers the  
appearance of one particular type of frozen density fluctuation,  
whereas the latter is directly related to the non-linear compressibility, i.e. the response to a pressure  
field that couples to the density. Other non-linear responses to 
a field that couples to the degrees of  
freedom undergoing a glass transition are also relevant (for example, the dielectric response when the dipoles are strongly  
coupled to the translational degrees of freedom, such as in glycerol, OTP, etc.). 
 
Let us directly focus on the dynamical susceptibility and postpone the discussion of its static limit 
to section II.B.3. 
Indeed, the analogy with spin-glasses has to be taken with a grain of salt (see section II.B.3).
The dynamical four-point density function defined as: 
\begin{equation}\label{chi4}   
G_4 (r,t) =  \langle  
\delta \rho_x(t=0)\delta \rho_x(t)\delta \rho_{x+r}(t=0)\delta\rho_{x+r}(t) \rangle - C^2(t)  
\qquad  
C(t) \equiv \langle\delta \rho_x(t=0)\delta \rho_x(t) \rangle,  
\end{equation}   
is related to the dynamical non-linear response  
of the system to an external excitation that couples to the density
\footnote{For practical reasons one has to introduce an overlap function 
in the definition of $G_4$ or focus on slightly different observables \cite{Glotzer,Berthier1}.}.  
Once again, the idea is that the change of the two point correlation between $x$ and $y$ induced by the external  
field of frequency $\omega \sim \tau^{-1}$ will be large if on that time scale, the dynamics at these two points is  
strongly correlated, which is true precisely if $G_4(x-y,\tau)$ is large. 
The extended, non-linear {\sc fdt} discussed  
in the next section makes this statement more precise and finally leads to our central results,  
Eqs. (\ref{result},\ref{efdt})   
\footnote{A related idea, explored in \cite{BBB}, is that the sensitivity   
of the two-body correlation function $C(t)$ to {\it temperature}   
changes is again related to four-point correlation functions, through  
a mechanism related to the one discussed in this paper.}. Now, recent numerical  
\cite{Harrowell,Onuki,Glotzer,Berthier1} and theoretical work \cite{FP,WBG,GC,BB1,TWBBB} have focused on the  
above choice of four point density function. Its integral over space $\chi_4(t)=\int d^dr G_4 (r,t)$   
(divided by $V$) gives the variance of the correlation function $C(t)$ for a system of finite volume $V$ \cite{Ludo},  
and is therefore a good quantitative measure for dynamical heterogeneities. This quantity  
was unambiguously shown to display a peak at $t = \tau$, of increasing amplitude as the temperature is   
decreased and the glass temperature is approached \cite{Harrowell,Onuki,Glotzer,Berthier1}, signaling   
increased cooperativity in the dynamics and the growth of a length scale $\ell$, which should in turn  
show up in the non-linear response of the system. 
  
\subsubsection{Different scenarii for the glass transition: qualitative predictions}  
  
 We therefore expect, on very general grounds, the non-linear  
 response to a field that couples to degrees of   
 freedom undergoing a collective freezing phenomenon, to  
 increase substantially (as $\ell^{2-\overline{\eta}}$)    
 as the glass phase is approached. However, as we discuss now, the  
 details of this increase do depend on the   
 specific scenario at play. Most important in that respect is the quantitative  
 relation between the cooperative length scale $\ell$   
 and the relaxation time $\tau$, which is often a power-law $\tau  
 \sim \ell^z$ where $z$ is the dynamical exponent.  
  
 One scenario for the glass state is based on the idea that some  
 mobility defects are needed to trigger the dynamics, which   
 slows down  at low temperatures because these defects become rare  
 \cite{SR,GC,BG}. Kinetically constrained models   
 provide an interesting framework to quantify this idea. The class  
 of so-called ``East'' models seems able to capture some   
 of the phenomenology of fragile glasses and predict a temperature  
 dependent exponent $z = z_0/T$, which may become   
 large at low temperature \cite{Sollich,BG}. This is very important  
 since the relaxation time of the system is known to   
 increase by 15 orders of magnitude as the temperature is reduced  
 from $1.5 T_g$ to $T_g$. But if, say, $z=15$  
 the cooperative length $\ell$ would only increase by a very moderate  
 factor $10$.   
  
 Another well known scenario is based on the   
 Mode-Coupling theory ({\sc mct}) of supercooled liquids, which  
 predicts a dynamical singularity  at a finite   
 temperature $T_{MCT}$, where the relaxation time should diverge as  
 $\tau \sim (T-T_{MCT})^{-\gamma}$ \cite{Gotze}, with  
 a non-universal exponent $\gamma$. It was recently   
 shown that this singularity is actually accompanied by the  
 divergence of a cooperative length $\ell$, precisely defined   
 in terms of the four-density correlation function above, Eq. (\ref{chi4}) \cite{BB1}.
 The nature of the transition is actually equivalent to that of the mean-field $p-$spin glass, 
 where both the spin-glass and the non-linear susceptibility diverge at $T^-_{MCT}$ \cite{FP}
 \footnote{To derive this result, one should compute $\chi_3$ within one of the states that
 dominates the partition function at temperature $T < T_{MCT}$.}.
 The exponent $z$ is found to be equal to $2 \gamma$;  
 reasonable values of $z$ being in the range $4-6$. However, the  
 ideal {\sc mct} phase transition is `avoided' in real   
 systems. Only the first $2-4$ decades of increase of $\tau$ can be  
 satisfactorily accounted by {\sc mct}, before some   
 new physics come  into play, that smear out the {\sc mct}  
 transition. In the temperature region where    
 $T < T_{MCT}$, the system should, according to mean-field, be  
 completely frozen. In finite dimensions, however,   
 barriers to motion are finite and the dynamics is instead strongly  
 activated. More precisely,   
 the liquid is a `mosaic' of local metastable glass states,  
 that must rearrange collectively \cite{AG,KTW,BB2}.   
 The size of these frozen clusters is the cooperative length $\ell$,  
 which increases as $T$ decreases, but now only logarithmically with $\tau$.  
 Therefore, within the Random First Order Theory  
({\sc rfot}) of \cite{KTW} which unifies 
 {\sc mct} and the mosaic scenario one expects a cross-over between a   
 power-law increase of $\ell$ for $T \sim T_{MCT}$ and a much more  
 modest increase of $\ell$ as the temperature is reduced  
 from $T_{MCT}$ to $T_g$. [Below $T_g$, aging effects come into play  
 and we expect that an equation similar to  
 Eq. (\ref{resultw}) will hold in that regime.]  
 Finally, the `avoided critical point' scenario of  
 Kivelson and Tarjus also predicts   
 a cooperative length that grows weakly (logarithmically) with the  
 relaxation time \cite{Tarjus}.   
  
 The value of $\overline{\eta}$ in Eq. (\ref{result}) above is not  
 known either, and presumably depends both   
 on the scenario and on the  temperature regime. Negative values of  
 $\overline{\eta} \approx - 1.58$ have been reported for East models \cite{BG},    
 whereas $\overline{\eta}$ is probably small in the  
 Mode-Coupling region \cite{BB1}. In the simplest   
 mosaic state scenario where clusters are compact,  the exponent  
 $2-\overline{\eta}$ is equal to the dimension   
 of space $d$. Rather independently of the scenario, we therefore  
 expect a noticeable rise of the non-linear susceptibility in    
 supercooled liquids as the temperature is reduced: taking  
 $\overline{\eta}=0$ and $\ell(T_g)/\ell(1.5 T_g) = 5$    
 leads to an increase of the peak value of $\chi_3$ by a factor  
 $\approx 25$. (Note however that depending on the particular physical
 observable there might be other contributions coming from the temperature dependence of   
 $\chi_s$ or, for example for the non-linear dielectric susceptibility, from the Lorentz field effects 
 that may play an important role in strongly dielectric systems \cite{LL}).    
 
\subsubsection{Glasses vs. spin-glasses: some caveats} 
 
The tricky aspect of the analogy between glasses and spin-glasses is that the {\it static} non-linear 
susceptibility of glasses is in fact not expected to display any divergence. As a matter of fact, no growing 
correlation length has ever yet been found in any static correlation function close to the glass transition, 
neither in experiments 
nor in simulations. This is not only because the glass transition is never sharp 
in real glasses, but also because from a purely theoretical point of view the situation is more subtle.  
Within {\sc rfot}, for example, the static non-linear susceptibility does {\it not} diverge.
The reason is that {\sc rfot} predicts an exponentially large number of possible amorphous states 
and thus a non zero configurational entropy for $T_K < T < T_{MCT}$ \cite{W5,MP,Cardenas} (where $T_K$  
is the `entropy crisis' temperature). Now, since equilibrium  
thermal averages are sums over all states, one has $\langle \delta \rho_x  \delta \rho_y \rangle  
\equiv \sum_{\alpha} w_\alpha \langle \delta \rho_x \delta \rho_y \rangle_{\alpha}$, where  
the subscript $\alpha$ indicates that the average is restricted to the metastable state $\alpha$, and  
$w_{\alpha}$ is the weight of state $\alpha$. Therefore: 
\be 
\sum_{y} \langle \delta \rho_x  \delta \rho_y \rangle^2 \equiv   
\sum_{\alpha,\beta}\sum_{y} w_{\alpha} w_{\beta}\langle \delta \rho_x   
\delta \rho_y \rangle_{\alpha}\langle \delta \rho_x  \delta \rho_y \rangle_{\beta}. 
\ee 
The divergence of the static spin-glass susceptibility is due to the diagonal terms $\alpha=\beta$.  
In the case of spin-glasses, the number of relevant states is effectively finite, and the above 
sum diverges in the spin-glass phase. Within {\sc rfot}, on the other hand, the number of  
relevant metastable states is so huge for $T_K < T < T_{MCT}$ that the diagonal contribution  
tends to zero, as does the 
off-diagonal contribution since the different frozen patterns are uncorrelated with each other. 
The only way to unveil any growing correlation is to focus on the terms $\alpha = \beta$, a 
calculation that is possible in mean-field (see previous footnote). In non mean-field 
situations, this can be done in two ways: one can study (a) the static correlations but in  
a self-induced static pinning field (freezing all the particles outside a cavity and 
studying the thermodynamics inside \cite{BB2}), which select a particular state $\alpha$ or  
(b) the dynamical correlations on a time scale short enough for the system to remain  
in a single state \cite{FP}. 
For example, the four-point correlation function defined in Eq. (\ref{chi4}) 
for $t \stackrel{<}{\sim} \tau$, effectively reproduces, for $\tau$ large enough, the static sum  
restricted to $\alpha=\beta$ \cite{FP}, since $\tau$ is the time needed to evolve from state $\alpha$  
to any another state. More precisely,  
one sees that in this regime, $G_4$ can be written in a form closer to the corresponding 
expression for spin-glasses: 
\begin{equation}\label{chi4bis}   
G_4 (r,t) \approx  [ \langle \delta \rho_x(t') \delta \rho_{x+r}(t') \rangle_{\alpha}    
\langle \delta \rho_x(t'+t)\delta\rho_{x+r}(t'+t) \rangle_{\alpha} ]_{t'},   
\end{equation}   
where the disorder average $[...]$ in the case of spin-glasses is replaced by an average over  
time, $[...]_{t'}$, corresponding to different configurations of the self-induced disorder in glasses.  
Therefore, we expect that the analogy with static spin-glasses indeed makes sense for  
$t \stackrel{<}{\sim} \tau$. A practical consequence of this observation is that the scaling function  
${\cal H}(x)$ in Eq. (\ref{result}) should tend to zero at zero frequency for glasses, at variance with the 
spin-glass case where it remains finite at $T=T_g^+$ and for zero field.  

Even the spin-glass case turns out to be tricky since, as mentioned above, the scaling function ${\cal H}(x \to 0)$ 
in fact also goes to zero for $H \neq 0$ in the context of the full replica symmetry breaking solution \cite{DGMMZ,Young}. 
Nevertheless, our dynamical result Eq. (\ref{efdt}) suggests this sum rule will generically {\it not} hold at 
finite frequency whenever the four-point correlation function has a non trivial time dependence.  This statement should 
of course be checked explicitely for mean-field models with continuous replica symmetry breaking \cite{ABB}.
If indeed ${\cal H}(x \sim 1)$ is found to be non zero close to the {\sc at} line, the experimental study of the dynamical 
non linear susceptibility, predicted to diverge for $\omega \tau \sim 1$, would offer a direct way to prove or disprove 
the existence of an {\sc at} line in real systems (see \cite{Young2} for a recent discussion).

 \section{Theoretical analysis}   
  
 In the rest of this paper, we give some theoretical justifications  
 of our  
 central result, Eqs. (\ref{result},\ref{resultw}). We  will use the  
 Langevin  
 equation formalism for continuous spins \cite{Guilhem}, but our  
 result are  
 expected to hold more generally   
 (for example, if the continuous spins are replaced by interacting  
 particles   
 with Newtonian dynamics).  
  
 \subsection{Linear response}   
  
 We assume that the equation of motion of spin $s_i$ is given by:   
 \be   
 \partial_t s_i = - \partial_{s_i} H + \eta_i(t),   
 \ee   
 where $H$ is the Hamiltonian of the system, which we do not specify  
 explicitly. In the case of spin glasses it contains quenched  
 disorder and   
 possibly one body terms ensuring an Ising like character to the  
 spins $s_i$. The coupling to    
 an external, site dependent field $h_i(t)$ amounts to add to $H$  
 the sum over spins of $h_i(t) s_i$. 
 The Gaussian noise $\eta_i$ is as usual of zero mean, white in time  
 and decorrelated from spin to spin:   
 \be   
 \langle \eta_i(t_1) \eta_j(t_2) \rangle = 2 k_B T \delta(t_1-t_2) \delta_{i,j}.  
 \ee   
 Since the noise is Gaussian, one can establish the following  
 identity:   
 \be   
 \langle s_i(t_1) \eta_j(t_2) \rangle = k_B T \left \langle  
 \frac{\partial s_i(t_1)}{\partial h_j(t_2)} \right \rangle.   
 \ee   
 Let us first quickly re-establish the standard linear {\sc fdt}.   
 From the above equation  
 and the equation of motion,    
 the response of a spin to an earlier field is:   
 \be   
 \chi_{ij}(t_1,t_2) = \left \langle \frac{\partial  
 s_i(t_1)}{\partial h_j(t_2)} \right \rangle =    
 \frac{1}{k_B T} \left \langle s_i(t_1) [ \partial_{t_2} s_j +  
 \partial_{s_j} H(t_2) ] \right \rangle.   
 \ee   
 The averaging above assumes the system to be in equilibrium: we  
 average over all histories with initial conditions   
 appearing with the equilibrium Boltzmann weight. The second term in  
 the right hand side is zero since, for an arbitrary    
 observable $O(\{t_a\})$ that depends on times $t_a$, all posterior  
 to $t_2$, one has:   
 \be    
 \left \langle O(\{t_i\}) \partial_{s_j} H(t_2) \right \rangle  
 \equiv - k_B T    
 \int \prod_a ds(t_a) P[\{s(t_a)\}|s_2] O(\{s(t_a)\}) ds_2  
 \partial_{s_2} \exp[-H(s_2)/k_B T] = 0,    
 \ee   
 where $s_2=s(t_2)$ and the last equality holds because the last term is a total  
 derivative. Therefore, one finds the    
 usual {\sc fdt} relation:   
 \be   
 \chi_{ij}(t_1,t_2) = \frac{1}{k_B T} \frac{d}{dt_2} \langle  
 s_i(t_1) s_j(t_2) \rangle.   
 \ee   
 Integrating this quantity over $t_2$ with a constant field  
 $h_i(t_2)=h$ gives the static susceptibility    
 $\chi_s$, which, as is well known, is found to be the integral over  
 space of the two-body correlation    
 function. In the case of a static critical point where the  
 correlation length   
 $\xi$ diverges, one would have $\chi_s \sim \mu ^2 \xi^{2-\eta}/k_B  
 T$,    
 where $\eta$ is the standard critical exponent of the static  
 transition and $\mu$ the elementary magnetic moment. However, in  
 the case of glassy systems, the two point   
 function is not critical and one rather expects $\chi_s \sim \mu  
 \xi^d/k_B T$ where $\xi$ remains microscopic and does not grow appreciably 
lowering the temperature (or increasing the density). 
As emphasized in Section II,  
 one should in the case of amorphous systems rather focus on non  
 linear effects to observe some non trivial   
 behaviour.   
  
 \subsection{Non-linear response: the static limit}  
  
 As a consequence we want to extend the above calculation to the  
 response at time $t_1 > t_2$ to   
 three field `kicks' at times $t_2 > t_3 > t_4$. This is given by:   
 \be   
 \chi_{3,ijkl}(t_1,t_2,t_3,t_4) = \left \langle \frac{\partial^3  
 s_i(t_1)}{\partial h_j(t_2) \partial h_k(t_3)   
 \partial h_l(t_4)} \right \rangle = (k_B T)^{-3} \langle s_i(t_1)  
 \eta_j(t_2) \eta_k(t_3) \eta_l(t_4) \rangle.   
 \ee   
 Using three times the Langevin equation of motion, and once the  
 above trick to get rid of the final    
 $\partial_{s_l} H(t_4)$, we find the following general relation, involving four terms:   
 \bea\label{full}   
 (k_B T)^{3} \chi_{3,ijkl}(t_1,t_2,t_3,t_4) &=&  
 \frac{d^3}{dt_2dt_3dt_4} \langle s_i(t_1) s_j(t_2)   
 s_k(t_3) s_l(t_4) \rangle + \\ \nonumber   
 \frac{d^2}{dt_3 dt_4} \langle s_i(t_1) \partial_{s_j} H(t_2)  
 s_k(t_3) s_l(t_4) \rangle &+&    
 \frac{d^2}{dt_2 dt_4} \langle s_i(t_1) s_j(t_2) \partial_{s_k} H  
 (t_3) s_l(t_4) \rangle + \\   
 \frac{d}{dt_4} \langle s_i(t_1) \partial_{s_j} H(t_2)  
 \partial_{s_k} H (t_3) s_l(t_4) \rangle.   
 \eea   
 Let us first analyze the static limit of this expression.  
 From the above result, one can show in full generality    
 that the static non-linear susceptibility $\chi_{3s}=  
 \chi_3(\omega=0,T)$,  
 obtained by integrating over    
 all $t_2 > t_3 > t_4$ with a constant field $h_i(t)=h$ on all  
 sites, is given by:   
 \be   
 (k_B T)^{3} \chi_{3s} = \frac{1}{N} \sum_{ijkl} \langle s_i(t_1)  
 s_j(t_1) s_k(t_1) s_l(t_1) \rangle_c,   
 \ee   
 where the subscript $c$ means that one takes the connected part of  
 the correlation and $N$ the total number    
 of sites. This result is exact and can be obtained directly using  
 equilibrium statistical mechanics. In the  
 present context, only the first term in expression Eq. (\ref{full})  
 for $\chi_{3,ijkl}(t_1,t_2,t_3,t_4)$   
 contributes for $\omega=0$. As discussed in Section II, the long  
 range order setting in spin-glasses is  
 unveiled not by the two-body correlation that oscillates in sign  
 and   
 averages to zero, but by the square of this two-body correlation.  
 Therefore, the  
 leading dominant term in the above sum corresponds to   
 the square of the two-body correlation obtained pairing $i,j,k,l$  
 --  
 say -- $i$ with $j$ and $k$ with $l$ within the two-body  
 correlation length   
 $\xi$ (which typically remains small at all temperatures):   
 \be   
 (k_B T)^{3} \chi_{3s} \approx - \frac{3}{N} \xi^{2d} \sum_{ik}  
 G_{ik}\qquad G_{ik}=\langle s_i(t_1) s_k(t_1) \rangle_c^2.   
 \ee   
 If one now assumes that $G_{ik}$ scales as in usual critical  
 phenomena \cite{FH,DGMMZ}:   
 \be   
 G_{ik} = \frac{1}{|r_i - r_k|^{d-2+\overline{\eta}}} {\cal  
 G}\left(\frac{|r_i-r_k|}{\ell}\right),   
 \ee   
 then the sum over $i,k$ behaves as $N \ell^{2 - \overline{\eta}}$,  
 finally leading to a static non-linear   
 susceptibility given by:   
 \be   
 \chi_{3s} \approx \frac{C \mu ^4 \xi^{2d}}{(k_B T)^{3}} \ell^{2 -  
 \overline{\eta}} \sim   
 \frac{\chi_s^2}{k_B T} \ell^{2 - \overline{\eta}}    
 \ee   
 where $C$ is a constant, and $\ell$ is counted in units of the  
 static correlation length $\xi$. This is the zero frequency result given in Eq. (\ref{result}).
 Note that for spin-glasses in a non zero field, a full replica symmetry breaking calculation 
 reveals that $G_{ik}=\langle s_i(t_1) s_k(t_1) \rangle_c^2$ is short-ranged \cite{DGMMZ}, 
 which means that $\chi_s$ is in fact non divergent.   
  
 \subsection{Non-linear response: dynamics}  
  
 The extension to non zero frequency of the above result  
 can proceed in different ways. Our result Eq. (\ref{result}) can be simply seen as a  
 standard dynamical scaling assumption close  
 to a critical point, as is indeed correct for spin-glasses  
 \cite{Zippelius,ParisiRanieri}. This result is expected to hold   
 whenever a critical point is responsible for the simultaneous increase  
 of the relaxation time and the cooperative length. This is true of   
 the Mode-Coupling Theory of glasses  
 \cite{FP,BB1}, and also of other scenarii discussed in the   
 introduction and in Section II, which rely on the existence of an  
 underlying critical point \cite{KTW,Sethna,Tarjus,MP,BB2,WBG}.   
 From a more technical point of view we want to justify  
 that the behaviour of the non-linear cubic response is the same as of  
 the first term on the right hand side of   
 Eq. (\ref{full}), whereas the three other terms are either  
 negligible or of the same order of magnitude (on   
 frequencies of the order of $\tau^{-1}$), but not more divergent.  
  
 A simple case that can be treated in some generality is when the  
 fluctuation of the norm of the spins   
 can be neglected, for example for Ising spins that can be recovered  
 from the Langevin equation   
 in the limit of infinitely sharp double well potential that is zero  
 if $s^{2}=1$ and infinite otherwise.  
 After several integration by parts and using $s^{2}=1$, one can  
 show that the three last terms of   
 Eq. (\ref{full}) do not contribute to the non-linear a. c.  
 susceptibility at low frequencies (much smaller than the microscopic, high 
 frequency scale of the model). 
 One is therefore left with the first term of Eq. (\ref{full}), that  
 contains three derivatives with   
 respect to time. If one assumes that the four-body correlation  
 $\langle s_i(t_1) s_i(t_2) s_k(t_3) s_k(t_4) \rangle$ is, for $|i-k| \sim \ell$, only a  
 function of $(t_1-t_2)/\tau$, $(t_2-t_3)/\tau$ and   
 $(t_3-t_4)/\tau$, the integration over $t_2,t_3,t_4$ with an  
 oscillating field at frequency $\omega$ and over space directly leads to Eq. (\ref{result}), i.e. a non-linear  
 susceptibility that scales as a certain function ${\cal H}$ of $\omega \tau$. This result is only justified in the 
 low frequency domain; for high frequency, contributions from the  short-time $\beta$-regime will obviously come into play. 
 Note that very generally, we expect ${\cal H}$ to be non trivial, although it does vanish at zero frequency whenever
 the static susceptibility is finite, as is the case for glasses and spin-glasses in an external field (see the 
 discussion in II.A.2, II.B.3).
  
 More generally, one can argue both physically and diagrammatically that the three last terms of Eq.  
 (\ref{full}) give contributions which are at  
 most of the same order of magnitude as the first one. From a  
 physical point of view, these terms contain less time derivatives  
 that the first, but also contain the  
 local `force' acting on the configuration, $\partial_{s} H(t)$.  
 Since we are interested in the low frequency   
 response of the system, we can decompose the dynamics of the spins  
 into a fast part $s^f$ and a slow part $s^*$, that   
 corresponds to the dynamics on a time of order $\tau$. It is clear  
 that the force acting on the slow modes can only   
 lead to a slow dynamics of these modes, i.e. $|\partial_{s^*} H|  
  \stackrel{<}{\sim}  \tau^{-1}$. Therefore, for frequencies $\sim \tau^{-1}$,  
 one has, for example,   
 \be  
\langle s_i(t_1) \partial_{s_j} H(t_2)  
 s_k(t_3) s_l(t_4) \rangle  \stackrel{<}{\sim}   \tau^{-1} 
 F\left(\frac{t_1-t_2}{\tau},\frac{t_2-t_3}{\tau},\frac{t_3-t_4}{\tau}\right),  
 \ee  
 (where $F$ is a certain function), which after integration leads again to a result of the form  
 (\ref{result}).  
  
 \begin{figure}  
 \epsfig{file=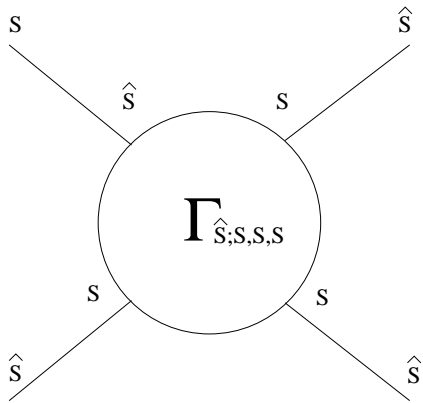,width=5.5cm}  
 \caption{\label{fig1} General diagrammatic representation of the  
 non-linear cubic response.}  
 \end{figure}   
  
 One can understand this result from a different point of view  
 using diagrams for a general Langevin equation, which leads to a   
 dynamical field theory with the spin field $s$   
 and the response field $\hat s$ \cite{Cardy}. The non-linear  
 cubic response $\chi_{3}$ at time $t_{1}$ to three instantaneous  
 fields at times  
 $t_{2},t_{3},t_{4}$ can be written in full generality as (see Fig.  
 \ref{fig1}):  
 \begin{equation}\label{gammadef}  
 \chi_{3}(t_{1};t_{2},t_{3},t_{4})=\int dt_{1}'dt_{2}'dt_{3}'dt_{4}'  
 \Gamma_{\hat s,  
 s,s,s}(t_{1}';t_{2}',t_{3}',t_{4}')\chi(t_{1}-t_{1}')\chi(t_{2}'-t_{2})\chi(t_{3}'-t_{3})  
 \chi(t_{4}-t_{4}')  
 \end{equation}  
 where $\Gamma_{\hat s,s,s,s}$ is the amputated vertex with legs  
 $\hat s,s,s,s$ (for simplicity we skip here the   
 space indices). Note that the vertex $\Gamma_{s,s,s,s}$ is zero by  
 causality because  
 it contains for sure a closed loop of response functions. The other  
 vertices do not appear because the correlation function $\langle  
 \hat s \hat s \rangle $ vanishes by causality.   
  
\begin{figure}  
 \epsfig{file=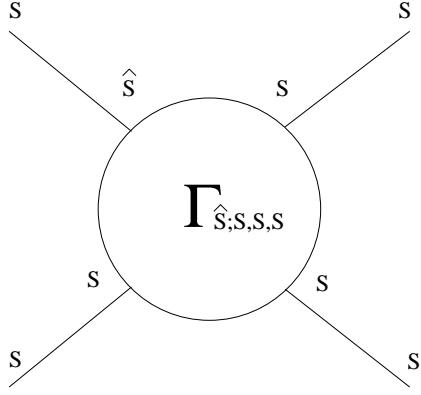,width=5.5cm}  
 \caption{\label{fig2} Diagrammatic representation of $G_4^{a}$}  
 \end{figure}

 Now let us consider the diagrams contributing to the connected four  
 body correlation function. There is   
 a first contribution $G_4^{a}$ obtained by plugging three two body  
 correlation functions  
 into $\Gamma_{\hat s,s,s,s}$, see Fig. \ref{fig2}.  It is  
 straightforward to check, using {\sc fdt}, that this   
 first series of diagrams, $G_4^{a}$, is directly related to the  
 non-linear cubic response. If $G_4^{a}$ was the   
 only contribution, then one would find an extended {\sc fdt} where  
 only the first term on the right hand side of   
 Eq. (\ref{full}) contributes.    
  
\begin{figure}  
 \epsfig{file=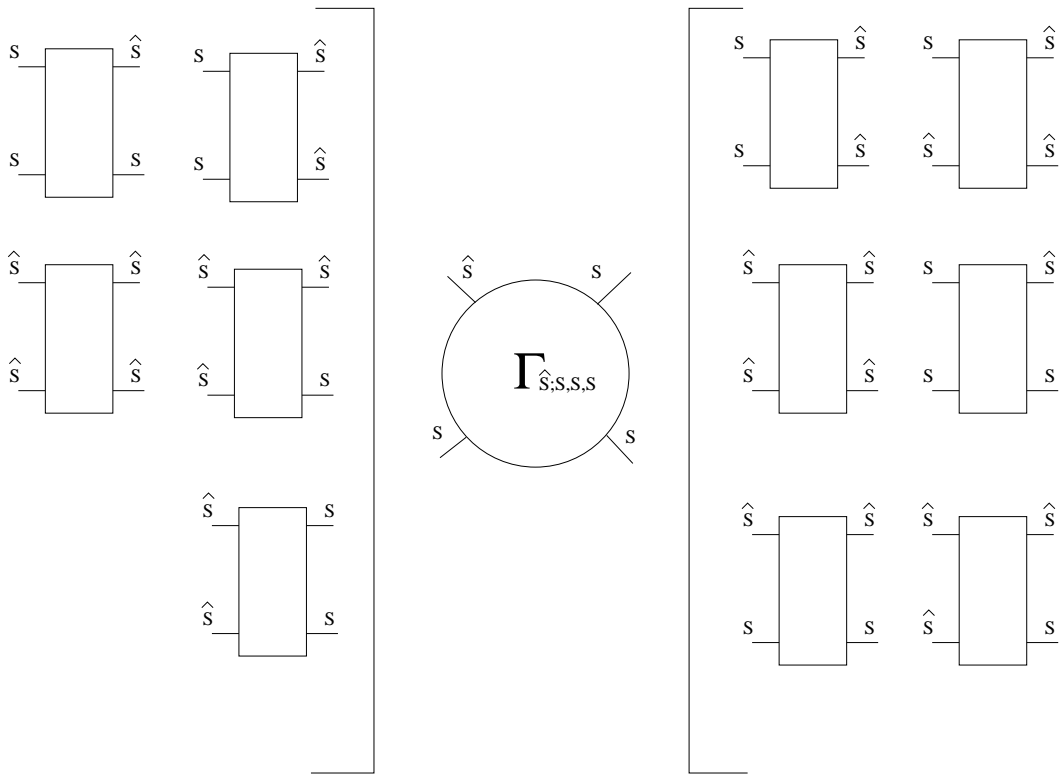,width=14cm}  
 \caption{\label{fig3} Diagrammatic representation of $G_4^{b}$}  
 \end{figure}

 There is another contribution, $G_4^{b}$,   
 that corresponds to constructing ladders with the irreducible  
 vertices $\Gamma_{s,s;\hat s,\hat s}^{irr},\Gamma _{s,s;s,\hat  
 s}^{irr}, \Gamma _{\hat s,\hat s;s,\hat s}^{irr}, \Gamma _{\hat  
 s,\hat  
 s;\hat s,\hat s}^{irr},  \Gamma _{\hat s,\hat  
 s;s,s}^{irr}$ on the left of $\Gamma _{\hat s,s,s,s}$ and   
 the irreducible vertices $\Gamma _{s,s;\hat s,\hat s}^{irr},\Gamma  
 _{s,\hat s;\hat s,\hat  
 s}^{irr}, \Gamma _{\hat s,\hat s;\hat  s,\hat s}^{irr}, \Gamma  
 _{\hat s,\hat s;s,\hat s}^{irr},    
 \Gamma _{\hat s,s;s,\hat s}^{irr},\Gamma _{s,s;s,\hat s}^{irr},$  
 on the right of $\Gamma _{\hat s,s,s,s}$. See fig. \ref{fig3}.  
 [We recall that the irreducible vertex $\Gamma_{1,2;3,4}^{irr}$ is  
 the sum  
 of all Feynman diagrams contributing to  $\Gamma_{1,2;3,4}$ (the  
 amputated vertex) that has the property that cutting two internal  
 lines does not   
 separate the diagram into two disconnected parts, such that  
 one part contains the lines  
 $1,2$ and the other one the lines $3,4$.] 
 Finally, the last contribution, $G_4^{c}$, is formed by plugging  
 together the same irreducible diagrams used in $G_4^{b}$ but  
 without making use of $\Gamma_{\hat s,s,s,s}$.   
  
 In the case of the critical equilibrium dynamics of spin-glasses,  
 Eq. (\ref{result}) is already known  
 \cite{Zippelius,ParisiRanieri} and could have been guessed   
 a priori from the general scaling properties of second order phase  
 transitions with a   
 single diverging length (and time) scale. As a consequence in this  
 case  
 $G_4^{b},G_4^{c}$ are of the same order or less divergent than  
 $G_4^{a}$. The case of structural glasses is a priori more tricky,  
 since there is no consensus   
 on the effective critical microscopic model that would describe  
 them. However, if we take as an  
 established fact (at least numerically) that the four body  
 correlation is governed by a length scale that  
 increases as the glass is approached, then this effect has to be  
 contained in (at least) one of the three   
 contributions $G_4^{a},G_4^{b},G_4^{c}$. Now, the non-linear  
 response certainly contains   
 the contribution related to $G_4^{a}$; therefore both $\chi_3$ and  
 $G_4$ grow (or even diverge) similarly   
 unless another family of (more) diverging diagrams (the ones  
 contributing to $G_4^{b},G_4^{c}$) can be   
 constructed. We believe that this is a rather unlikely scenario and  
 instead  
 we expect that in general $G_4^{a}$ and $G_4^b$ are of the same  
 order of   
 magnitude, and $G_4^c$ is sub-dominant, in which case Eq.  
 (\ref{result}) holds.   
 Strictly speaking, these arguments prove that if the   
 non-linear cubic dynamical response diverges, a similar (or stronger) 
divergence  
 is   
 expected for the four body correlation function, but not  
 vice-versa.   
 Therefore it would be important to check our prediction for  
 specific models of the   
 glass transition, in the spirit of \cite{TWBBB}. Here, we just want  
 to emphasize that   
 the techniques used in \cite{BB1} can be used to establish that our  
 central result, Eq. (\ref{result}),   
 holds within the Mode-Coupling Theory of the glass transition.  
  
 Finally, let us remark that the extension to the non-equilibrium  
 case  
 can be tackled in a similar way. In particular, since the four-body  
 correlation function diverges with $t_{w}$   
 \cite{ChamonCugliandolo} in spin-glasses, and the classification in  
 terms of $G_4^{a},G_4^{b},G_4^{c}$ carries over   
 to the non-equilibrium case, the above discussion can be  
 generalized to the non-equilibrium case as well.   
  
 In summary, we have shown in this section that for glassy systems  
 close to a critical point, where the    
 relaxation time and cooperative length diverge, an extended  
 approximate {\sc fdt} relates the non-linear   
 susceptibility to the four-point correlation function in the low frequency domain:   
 \be\label{efdt}   
 (k_B T)^{3} \chi_{3,ijkl}(t_1,t_2,t_3,t_4)\sim  
 \frac{d^3}{dt_2dt_3dt_4} \langle s_i(t_1) s_j(t_2)   
 s_k(t_3) s_l(t_4) \rangle;   
 \ee  
 where $\sim$ means that right and left hand side have the same  
 critical behaviour.  
 The additional terms missing in the above equation are either of  
 the same order of magnitude, or  
 negligible.   
  
 \section{Conclusion}  
  
 In conclusion, we have shown in this paper that if the abrupt  
 slowing down of glassy materials is accompanied by the   
 growth of a cooperative length $\ell$, then the non-linear, 
 $3 \omega$ response to an oscillating field 
 (at frequency $\omega$) should    
 substantially increase and give precious information on the  
 temperature (or density) dependence of $\ell$.    
 The theoretical motivation is that the non-linear susceptibility is  
 approximatively related, for glassy systems close   
 to a critical point, to the four-point correlation function that  
 captures dynamical cooperativity. This relation is  
 certainly correct within the context of the Mode-Coupling Theory of glasses, but should hold in  
 other cases as well.
  
 In supercooled liquids, the analysis of the non-linear  
 compressibility (sound wave harmonics) should allow one to probe  
 directly the existence of a growing cooperative length. This should also 
 be true of the non-linear dielectric susceptibility, at least in systems where the dipoles are  
 strongly coupled to the glassy degrees of freedom.   
 Although early experiments seemed to show no interesting effects  
 \cite{Wu}, we believe that more systematic studies   
 should be performed \cite{LL}, especially now that numerical  
 simulations have unambiguously shown the growth   
 of a cooperative length in the four-point function  
 \cite{Harrowell,Onuki,Glotzer,Berthier1,WBG}.   
 These experiments should also allow one to bridge the gap between  
 the length-scales observed on simulation time scales and  
 the length-scales observed experimentally on much larger time-scales  
 close to the glass transition temperature \cite{Ediger}.  
 The study of non-linear specific heat effects, although more  
 complex, may be interesting too \cite{Nagel,All}.   
 From a more general point of view any non-linear dynamical response  
 (for example, non-linear rheology in soft glassy  
 materials) should be worth studying if the corresponding linear  
 response can be used as a probe of slow dynamics.  
  
 In spin-glasses, non-linear a.c. magnetic susceptibility measurements in non-zero
 field could shed light on the existence of a de Almeida-Thouless line. In the 
 aging phase, such measurements should allow one to test in more details the length scale ideas put forward in  
 \cite{FH,BM,KH,Bou,Ghost,LudoKT}. Compared to the case of glasses, the experimental situation is particularly  
 encouraging since the non-linear susceptibility is already known to diverge at the spin-glass transition. 
 There should be a clear trace of this divergence in the aging phase, except if
 some subtle cancellation occurs even at non zero frequency (the mean-field replica theory indeed predicts 
 such a cancellation in the static case). The effect of temperature cycling on the non-linear susceptibility should 
 then give direct indications of the mechanisms of rejuvenation and memory \cite{Bou,Ghost}. We therefore hope that 
 the ideas expressed in this paper will help shed light on the issue of    
 dynamical heterogeneity and cooperativity in disordered, amorphous systems. 
  
 \begin{acknowledgments}   
  
 We thank L. Berthier, A. Billoire, A. Bray, C. De Dominicis, V. Dupuis, M. M\'ezard, M. Moore, E. Vincent 
 and M. Wyart for discussions. We are 
 grateful to F. Ladieu, D. L'Hote for important comments, in particular on Lorentz field effects, 
 and for the interest and enthusiasm with which they started experiments along the lines of this study. 
 We also thank P. Young for a crucial remark on the case of a spin-glass in a field, and L. Borland for 
 carefully reading the manuscript. G. B. is partially supported by the European Community's Human Potential Programme
 contracts HPRN-CT-2002-00307 (DYGLAGEMEM). 
  
\end{acknowledgments}

\end{document}